\documentclass[12pt]{article}

\usepackage{amsmath,amssymb}

\def\a{\alpha'}

\begin{document}

\hyphenation{chi-ra-li-ty}

\hyphenation{cor-res-pond}

\hyphenation{li-ne-ari-zed}

\hyphenation{jeo-par-dy}

\hyphenation{dif-fe-ren-tial}

\hyphenation{re-pre-sent}

\hyphenation{res-tric-tive}

\begin{titlepage}
\begin{center}

\vskip 20mm

{\Huge One loop superstring effective actions and ${\cal N}=8$
supergravity}

\vskip 10mm

Filipe Moura

\vskip 4mm

{\em Security and Quantum Information Group - Instituto de
Telecomunica\c c\~oes\\ Instituto Superior T\'ecnico, Departamento
de Matem\'atica \\Av. Rovisco Pais, 1049-001 Lisboa, Portugal}

\vskip 4mm

{\tt fmoura@math.ist.utl.pt}

\vskip 6mm

\end{center}

\vskip .2in

\begin{center} {\bf Abstract } \end{center}
\begin{quotation}\noindent

In a previous article we have shown the existence of a new
independent ${\cal R}^4$ term, at one loop, in the type IIA and
heterotic effective actions, after reduction to four dimensions,
besides the usual square of the Bel-Robinson tensor. It had been
shown that such a term could not be directly supersymmetrized, but
we showed that was possible after coupling to a scalar chiral
multiplet.

In this article we study the extended (${\cal N}=8$)
supersymmetrization of this term, where no other coupling can be
taken. We show that such supersymmetrization cannot be achieved at
the linearized level. This is in conflict with the theory one gets
after toroidal compactification of type II superstrings being ${\cal
N}=8$ supersymmetric. We interpret this result in face of the recent
claim that perturbative supergravity cannot be decoupled from string
theory in $d \geq 4$, and ${\cal N}=8$, $d=4$ supergravity is in the
swampland.
\end{quotation}

\vfill


\end{titlepage}

\eject

\tableofcontents

\section{Introduction}
\setcounter{equation}{0} \indent

String theories require higher order in $\a$ corrections to their
corresponding low energy supergravity effective actions. Among these
corrections, at order $\a^3$, the ${\cal R}^4$ terms (the fourth
power of the Riemann tensor) are present in type II
\cite{Gross:1986iv, Grisaru:1986px} and heterotic
\cite{Gross:1986mw} superstrings and in M-theory
\cite{Green:1997as}. These corrections need to be supersymmetric;
the topic of their supersymmetrization has been object of research
for a long time \cite{Peeters:2000qj, deRoo:1992zp}.

These corrections are also present in four dimensional supergravity
theories. Originally they were looked as candidate counterterms to
these theories, which were believed to be divergent. From the string
theory point of view they are seen as compactified string
corrections. In any case these corrections must be supersymmetric.
The number $\mathcal{N}$ of four-dimensional supersymmetries and
different matter couplings depend crucially on the manifold where
the compactification is taken.

The four-dimensional supersymmetrization of ${\cal R}^4$ terms has
been considered both in simple \cite{Deser:1977nt, Moura:2001xx,
Moura:2002ft} and in extended \cite{Deser:1978br, Kallosh:1980fi, Howe:1980th,
Moura:2002ip} supergravities. Although there are two
independent ${\cal R}^4$ terms in $d=4$, all these cases only
studied one such term: the square of the Bel-Robinson. Indeed, in
another article \cite{Christensen:1979qj} it is shown that the other
four-dimensional ${\cal R}^4$ term is part of a class of terms which
are not supersymmetrizable.

That term has not deserved any further attention until recently.
In our previous paper \cite{Moura:2007ks}, we have computed the
dimensional reduction, to four dimensions, on a torus, of the
ten-dimensional ${\cal R}^4$ terms from type II and heterotic
superstrings.\footnote{The ${\cal R}^4$ term from M-theory, when
reduced to $d=10$ on ${\mathbb S}^1$, results on the one-loop
${\cal R}^4$ term in type IIA superstring. The results of
\cite{Moura:2007ks} therefore also include the toroidal
compactification of M-theory to $d=4$.} We have then shown that
the other ${\cal R}^4$ term is part of the heterotic and type IIA
effective actions, at one loop, when compactified to $d=4$. Now,
when compactified to $d=4$ on a 6-torus ${\mathbb T}^6$, should be
respectively ${\mathcal N}=4, 8$ supersymmetric. Plus, ${\mathbb
T}^6$ is the most basic manifold one can think of in order to
compactify a ten-dimensional theory; all the terms one gets from
this compactification are present when one rather takes a more
complicated manifold. This means the new (or less known) ${\cal
R}^4$ term is present in any compactification to $d=4$ of type IIA
and heterotic superstrings.

In our previous work \cite{Moura:2007ks} we focused on ${\mathcal
N}=1$ supergravity. By taking a coupling to a chiral multiplet, we
were able to circumvent the argument of  \cite{Christensen:1979qj}
and indeed include the less known ${\cal R}^4$ term in an
${\mathcal N}=1$ supersymmetric lagrangian.

In this work we focus particularly on maximal ${\mathcal N}=8$
supergravity, the most restrictive of all the $d=4$ theories (its
multiplet is unique, and there are no matter couplings to take),
and one of the main reasons is precisely because this is the
theory which results after compactifying type IIA supergravity on
${\mathbb T}^6.$ Besides, the study of higher order corrections in
${\mathcal N}=8$ supergravity is particularly relevant considering
the recent claims that this theory may actually be eight-loop
finite \cite{Green:2006gt} or even ultraviolet finite
\cite{Bern:2006kd}.

In section 2 we will review and summarize some of the results of
\cite{Moura:2007ks}, concerning ${\cal R}^4$ terms in $d=10$ and
their reduction to $d=4.$ In section 3 we briefly review
linearized $d=4$ extended supergravity in superspace and some
known higher order linearized extended superinvariants and the
symmetries they should preserve. In section 4 we proceed with
trying to supersymmetrize in ${\mathcal N}=8$ the other less
known, but existing, ${\cal R}^4$ term using different
possibilities.


\section{${\cal R}^4$ terms in $d=10$ and $d=4$}
\setcounter{equation}{0} \indent


\subsection{${\cal R}^4$ terms in $d=10$}
\indent

The superstring $\a^3$ effective actions contain two independent
bosonic terms $I_X, I_Z,$ from which two separate superinvariants
are built \cite{Peeters:2000qj, Tseytlin:1995bi}. These terms are
given, at linear order in the NS-NS gauge field $B_{mn}$, by:
\begin{eqnarray}
I_X&=&t_8 t_8 {\cal R}^4 + \frac{1}{2} \varepsilon_{10} t_8 B
{\cal R}^4, \nonumber \\ I_Z&=&-\varepsilon_{10} \varepsilon_{10}
{\cal R}^4 +4 \varepsilon_{10} t_8 B {\cal R}^4. \label{ixiz}
\end{eqnarray}
For the heterotic string another two independent terms $Y_1$ and
$Y_2$ appear at order $\a^3$ \cite{Peeters:2000qj,
Tseytlin:1995bi, deRoo:1992zp}. Their parts which involve only the
Weyl tensor are given respectively by
\begin{eqnarray}
Y_1 &:=& t_8 \left(\mbox{tr} {\cal W}^2\right)^2, \nonumber \\ Y_2
&:=& t_8 \mbox{tr} {\cal W}^4 = \frac{X}{24} + \frac{Y_1}{4}.
\label{y1y2}
\end{eqnarray}
Each $t_8$ tensor has eight free spacetime indices. It acts in
four two-index antisymmetric tensors, as defined in
\cite{Gross:1986iv, Grisaru:1986px}. In our case,
\begin{eqnarray}
t_8 t_8 {\cal R}^4 &=&t^{mnpqrstu} t^{m' n' p' q' r' s' t' u'} {\cal
R}_{mn m' n'} {\cal R}_{pq p' q'} {\cal R}_{rsr' s'} {\cal R}_{tut'
u'}, \nonumber
\\ \varepsilon_{10} t_8 B {\cal R}^4 &=&t^{mnpqrstu} \varepsilon^{vw m' n'
p' q' r' s' t' u'} B_{vw} {\cal R}_{mn m' n'} {\cal
R}_{pq p' q'} {\cal R}_{rsr' s'} {\cal R}_{tut' u'}, \nonumber \\
\varepsilon_{10} \varepsilon_{10} {\cal R}^4 &=&
\varepsilon_{vw}^{\, \, \, \, \, \, \, \, mnpqrstu} \varepsilon^{vw
m' n' p' q' r' s' t' u'} {\cal R}_{mn m' n'} {\cal R}_{pq p' q'}
{\cal R}_{rsr' s'} {\cal R}_{tut' u'} \label{ixizi}.
\end{eqnarray}

The effective action of type IIB theory must be written, because of
its well known SL$(2,{\mathbb Z})$ invariance, as a product of a
single linear combination of order $\a^3$ invariants and an overall
function of the complexified coupling constant $\Omega= C^0 + i
e^{-\phi},$ $C^0$ being the axion. The order $\a^3$ part of this
effective action which involves only the Weyl tensor is given in the
string frame by
\begin{equation}
\left. \frac{1}{\sqrt{-g}} {\mathcal L}_{\mathrm{IIB}}
\right|_{\a^3} = -e^{-2 \phi} \a^3 \frac{\zeta(3)}{3 \times
2^{10}} \left(I_X - \frac{1}{8} I_Z \right) - \a^3 \frac{1}{3
\times 2^{16} \pi^5} \left(I_X - \frac{1}{8} I_Z \right).
\label{2bea}
\end{equation}
The corresponding part of the action of type IIA superstrings has
a relative "-" sign flip in the one loop term
\cite{Antoniadis:1997eg}. This  sign difference is because of the
different chirality properties of type IIA and type IIB theories,
which reflects on the relative GSO projection between the left and
right movers:
\begin{equation}
\left. \frac{1}{\sqrt{-g}} {\mathcal L}_{\mathrm{IIA}}
\right|_{\a^3} = -e^{-2 \phi} \a^3 \frac{\zeta(3)}{3 \times
2^{10}} \left(I_X - \frac{1}{8} I_Z \right) - \a^3 \frac{1}{3
\times 2^{16} \pi^5} \left(I_X + \frac{1}{8} I_Z \right).
\label{2aea}
\end{equation}
Heterotic string theories in $d=10$ have ${\mathcal N}=1$
supersymmetry, which allows corrections already at order $\a$,
including ${\mathcal R}^2$ corrections. These corrections come both
from three and four graviton scattering amplitudes and anomaly
cancellation terms (the Green-Schwarz mechanism). Up to order
$\a^3$, the terms from this effective action which involve only the
Weyl tensor are given in the string frame by
\begin{eqnarray}
\left. \frac{1}{\sqrt{-g}} {\mathcal L}_{\mathrm{heterotic}}
\right|_{\a+ \a^3} &=& e^{-2 \phi} \left[\frac{1}{16} \a \mbox{tr}
{\cal R}^2 +\frac{1}{2^9} \a^3 Y_1 - \frac{\zeta(3)}{3 \times
2^{10}} \a^3 \left(I_X - \frac{1}{8} I_Z \right) \right] \nonumber
\\ &-& \a^3 \frac{1}{3 \times 2^{14} \pi^5} \left(Y_1+ 4 Y_2
\right). \label{hea}
\end{eqnarray}

In order to consider these terms in the context of supergravity,
one should write them in the Einstein frame. To pass from the
string to the Einstein frame, we redefine the metric in $d$
(noncompact) dimensions through a conformal transformation
involving the dilaton, given by
\begin{eqnarray} g_{\mu\nu}
&\rightarrow& \exp \left( \frac{4}{d-2} \phi \right) g_{\mu\nu},
\nonumber \\ {{\cal R}_{\mu\nu}}^{\rho\sigma} &\rightarrow& \exp
\left(-\frac{4}{d-2} \phi \right) {\widetilde{{\cal
R}}_{\mu\nu}}^{\ \ \ \rho\sigma}, \label{rsre}
\end{eqnarray}
with ${\widetilde{{\cal R}}_{\mu\nu}}^{\ \ \ \rho\sigma}={{\cal
R}_{\mu\nu}}^{\rho\sigma} -
{\delta_{\left[\mu\right.}}^{\left[\rho\right.} \nabla_{\left.\nu
\right]} \nabla^{\left.\sigma \right]} \phi.$

Let $I_i({\cal R, M})$ be an arbitrary term in the string frame
lagrangian. $I_i({\cal R, M})$ is a function, with conformal
weight $w_i$, of any given order in $\a$, of the Riemann tensor
$\cal R$ and any other fields - gauge fields, scalars, and also
fermions - which we generically designate by $\cal M$. The
transformation above takes $I_i({\cal R,M})$ to $e^{\frac{4}{d-2}
w_i \phi} I_i({\cal \widetilde{R}, M}).$ After considering all the
dilaton couplings and the effect of the conformal transformation
on the metric determinant factor $\sqrt{-g},$ the string frame
lagrangian
\begin{equation} \label{esf} \frac{1}{2}
\sqrt{-g}\ \mbox{e}^{-2 \phi} \Big( -{\cal R} + 4 \left(
\partial^\mu \phi \right) \partial_\mu \phi + \sum_i I_i({\cal R,M}) \Big)
\end{equation} is converted into the Einstein frame lagrangian \begin{equation}
\label{eef} \frac{1}{2} \sqrt{-g} \left( -{\cal R} - \frac{4}{d-2}
\left( \partial^\mu \phi \right) \partial_\mu \phi + \sum_i
\mbox{e}^{\frac{4}{d-2} \left( 1 + w_i \right) \phi} I_i({\cal
\widetilde{R},M}) \right).
\end{equation}
Next we will take the terms we wrote above, but reduced to four
dimensions, in the Einstein frame.


\subsection{${\cal R}^4$ terms in $d=4$}
\indent

In four dimensions, the Weyl tensor can be decomposed in its
self-dual and antiself-dual parts \footnote{We used latin letters -
$m,n, \ldots$ - to represent ten dimensional spacetime indices. From
now on we will be only working with four dimensional spacetime
indices which, to avoid any confusion, we represent by greek letters
$\mu, \nu, \ldots$}:
\begin{equation}
{\cal W}_{\mu \nu \rho \sigma}= {\cal W}^+_{\mu \nu \rho \sigma} +
{\cal W}^-_{\mu \nu \rho \sigma}, {\cal W}^{\mp}_{\mu \nu \rho
\sigma} :=\frac{1}{2} \left({\cal W}_{\mu \nu \rho \sigma} \pm
\frac{i}{2} \varepsilon_{\mu \nu}^{\ \ \ \lambda \tau} {\cal
W}_{\lambda \tau \rho \sigma} \right). \label{wpm}
\end{equation}
The totally symmetric Bel-Robinson tensor is given in four
dimensions by ${\cal W}^+_{\mu \rho \nu \sigma} {\cal W}^{- \rho \
\sigma}_{\tau \ \lambda}.$ In the van der Warden notation, using
spinorial indices \cite{Penrose:1985jw}, to ${\cal W}^+_{\mu \rho
\nu \sigma},  {\cal W}^-_{\mu \rho \nu \sigma}$ correspond the
totally symmetric ${\cal W}_{ABCD}, {\cal W}_{\dot A \dot B \dot C
\dot D}$ being given by (in the notation of \cite{Moura:2002ft})
$${\cal W}_{ABCD}:=-\frac{1}{8} {\cal W}^+_{\mu \nu \rho \sigma}
\sigma^{\mu \nu}_{\underline{AB}} \sigma^{\rho
\sigma}_{\underline{CD}}, \, {\cal W}_{\dot A \dot B \dot C \dot
D}:=-\frac{1}{8} {\cal W}^-_{\mu \nu \rho \sigma} \sigma^{\mu
\nu}_{\underline{\dot A \dot B}} \sigma^{\rho
\sigma}_{\underline{\dot C \dot D}}.$$ The decomposition
(\ref{wpm}) is written as
\begin{equation}
{\cal W}_{A \dot A B \dot B C \dot C D \dot D}= -2
\varepsilon_{\dot A \dot B} \varepsilon_{\dot C \dot D} {\cal
W}_{ABCD}  -2 \varepsilon_{AB} \varepsilon_{CD} {\cal W}_{\dot A
\dot B \dot C \dot D}. \label{wpms}
\end{equation}
The Bel-Robinson tensor is
simply given by ${\cal W}_{ABCD} {\cal W}_{\dot A \dot B \dot C
\dot D}$.

In four dimensions, there are only two independent real scalar
polynomials made from four powers of the Weyl tensor
\cite{Fulling:1992vm}, given by
\begin{eqnarray}
{\cal W}_+^2 {\cal W}_-^2 &=& {\cal W}^{ABCD} {\cal W}_{ABCD}
{\cal W}^{\dot A \dot B \dot C \dot D} {\cal W}_{\dot A \dot B
\dot C \dot D}, \label{r441}\\ {\cal W}_+^4+{\cal W}_-^4 &=&
\left({\cal W}^{ABCD} {\cal W}_{ABCD}\right)^2 + \left({\cal
W}^{\dot A \dot B \dot C \dot D} {\cal W}_{\dot A \dot B \dot C
\dot D}\right)^2. \label{r442}
\end{eqnarray}
In particular, the Weyl-dependent parts of the invariants $I_X,
I_Z, Y_1, Y_2,$ when computed directly in four dimensions (i.e.
replacing the ten dimensional indices $m,n, \ldots$ by the four
dimensional indices $\mu, \nu, \ldots$), should be expressed in
terms of them. The details of the calculation can be seen in
\cite{Moura:2007ks}; the resulting ${\cal W}^4$ terms are
\begin{eqnarray}
I_X - \frac{1}{8} I_Z &=& 96 {\cal W}_+^2 {\cal W}_-^2, \\ I_X +
\frac{1}{8} I_Z &=& 48 \left({\cal W}_+^4 + {\cal W}_-^4 \right)
+672 {\cal W}_+^2 {\cal W}_-^2, \\ Y_1&=& 8 {\cal W}_+^2 {\cal
W}_-^2, \\Y_1 + 4 Y_2 = \frac{I_X}{6} + 2 Y_1 &=& 80 {\cal W}_+^2
{\cal W}_-^2 + 4 \left({\cal W}_+^4 + {\cal W}_-^4 \right).
\end{eqnarray}
$I_X - \frac{1}{8} I_Z$ is the only combination of $I_X$ and $I_Z$
which in $d=4$ does not contain (\ref{r442}), i.e. which contains
only the square of the Bel-Robinson tensor (\ref{r441}).
Interestingly, from (\ref{ixiz}) exactly this very same combination
is the only one which does not depend on the ten dimensional
$B^{mn}$ field and, therefore, due to its gauge invariance, is the
only one that can appear in string theory at arbitrary loop order.

We should consider another possibility: could there be any
four-dimensional ${\cal W}^4$ terms coming from the original
ten-dimensional $I_X+ \frac{1}{8} I_Z$ term in (\ref{ixiz}), but
this time including the (four dimensional) $B^{\mu\nu}$ field, as
a scalar, after toroidal compactification and dualisation (for a
detailed treatment see \cite{Cremmer:1997ct})? Let's take
\begin{equation}
\partial^{\left[\mu\right.} B^{\left.\nu \rho\right]}=
\epsilon^{\mu \nu \rho \sigma} \partial_{\sigma} D.
\end{equation}
$B^{\mu\nu}$ is a pseudo 2-form under parity; after dualisation in
$d=4,$ $D$ is a true scalar. This way, from the $\varepsilon_{10}
t_8 B {\cal R}^4$ term in $d=10$ one gets in $d=4,$ among other
terms, derivatives of scalars and at most an ${\cal R}^2$ factor.
(One also gets simply derivatives of scalars, without any Riemann
tensor.) An ${\cal R}^4$ factor would only come, after dualisation,
from a higher-order term, always multiplied by derivatives of
scalars. Therefore we cannot get any ${\cal R}^4$ terms this way.

We then write the effective actions (\ref{2bea}), (\ref{2aea}),
(\ref{hea}) in four dimensions, in the Einstein frame (considering
only terms which are simply powers of the Weyl tensor, without any
other fields except their couplings to the dilaton, and
introducing the $d=4$ gravitational coupling constant $\kappa$):
\begin{eqnarray}
\left. \frac{\kappa^2}{\sqrt{-g}} {\mathcal L}_{\mathrm{IIB}}
\right|_{{\cal R}^4} &=& - \frac{\zeta(3)}{32} e^{-6 \phi} \a^3
{\cal W}_+^2 {\cal W}_-^2 - \frac{1}{2^{11} \pi^5} e^{-4 \phi}\a^3
{\cal W}_+^2 {\cal W}_-^2, \label{2bea4} \\ \left.
\frac{\kappa^2}{\sqrt{-g}} {\mathcal L}_{\mathrm{IIA}}
\right|_{{\cal R}^4} &=& - \frac{\zeta(3)}{32} e^{-6 \phi} \a^3
{\cal W}_+^2 {\cal W}_-^2 \nonumber \\ &-& \frac{1}{2^{12} \pi^5}
e^{-4 \phi}\a^3 \left[\left({\cal W}_+^4 + {\cal W}_-^4 \right)
+224 {\cal W}_+^2 {\cal W}_-^2 \right], \label{2aea4}
\\ \left. \frac{\kappa^2}{\sqrt{-g}} {\mathcal L}_{\mathrm{het}}
\right|_{{\cal R}^2 + {\cal R}^4} &=& -\frac{1}{16} e^{-2 \phi} \a \left({\cal
W}_+^2 + {\cal W}_-^2 \right) +\frac{1}{64} \left(1-2 \zeta(3)
\right) e^{-6 \phi} \a^3 {\cal W}_+^2 {\cal W}_-^2 \nonumber \\
&-& \frac{1}{3\times2^{12} \pi^5} e^{-4 \phi}\a^3
\left[\left({\cal W}_+^4 + {\cal W}_-^4 \right) +20 {\cal W}_+^2
{\cal W}_-^2 \right]. \label{hea4}
\end{eqnarray}
These are only the moduli-independent ${\cal R}^4$ terms. Strictly
speaking not even these terms are moduli-independent, since they
are all multiplied by the volume of the compactification manifold,
a factor we omitted for simplicity. But they are always present,
no matter which compactification is taken. The complete action,
for every different compactification manifold, includes many other
moduli-dependent terms which we do not consider here: we are
mostly interested in a ${\mathbb T}^6$ compactification.


\subsection{${\cal R}^4$ terms and $d=4$ supersymmetry}
\indent

We are interested in the full supersymmetric completion of ${\cal
R}^4$ terms in $d=4.$ In general each superinvariant consists of a
leading bosonic term and its supersymmetric completion, given by a
series of terms with fermions.

The supersymmetrization of the square of the Bel-Robinson tensor
${\cal W}_+^2 {\cal W}_-^2$ has been known for a long time, in
simple \cite{Deser:1977nt, Moura:2001xx} and extended
\cite{Deser:1978br,Moura:2002ip} four dimensional supergravity.
For the term ${\cal W}_+^4 + {\cal W}_-^4$ there is a "no-go
theorem", which goes as follows \cite{Christensen:1979qj}: for a
polynomial $I({\cal W})$ of the Weyl tensor to be
supersymmetrizable, each one of its terms must contain equal
powers of ${\cal W}^+_{\mu \nu \rho \sigma}$ and ${\cal W}^-_{\mu
\nu \rho \sigma}$. The whole polynomial must then vanish when
either ${\cal W}^+_{\mu \nu \rho \sigma}$ or ${\cal W}^-_{\mu \nu
\rho \sigma}$ do.

The derivation of this result is based on ${\mathcal N}=1$ chirality
arguments, which require equal powers of the different chiralities
of the gravitino in each term of a superinvariant. The rest follows
from the supersymmetric completion. That is why the only exception
to this result is ${\cal W}^2 = {\cal W}_+^2 + {\cal W}_-^2$: in
$d=4$ this term is part of the Gauss-Bonnet topological invariant
(it can be made equal to it with suitable field redefinitions). This
term plays no role in the dynamics and it is automatically
supersymmetric; its supersymmetric completion is 0 and therefore
does not involve the gravitino.

The derivation of \cite{Christensen:1979qj} has been obtained
using ${\mathcal N}=1$ supergravity, whose supersymmetry algebra
is a subalgebra of ${\mathcal N}>1$. Therefore, it should remain
valid for extended supergravity too. But one must keep in mind the
assumptions which were made, namely the preservation by the
supersymmetry transformations of $R$-symmetry which, for
${\mathcal N}=1$, corresponds to U(1) and is equivalent to
chirality. In extended supergravity theories $R-$symmetry is a
global internal $\mbox{U}\left({\mathcal N}\right)$ symmetry,
which generalizes (and contains) U(1) from ${\mathcal N}=1$.

Preservation of chirality is true for pure ${\mathcal N}=1$
supergravity, but to this theory and to most of the extended
supergravity theories one may add matter couplings and extra terms
which violate U(1) $R$-symmetry and yet can be made
supersymmetric, inducing corrections to the supersymmetry
transformation laws which do not preserve U(1) $R$-symmetry.

That was the procedure taken in \cite{Moura:2007ks}, where the
${\mathcal N}=1$ supersymmetrization of (\ref{r442}) was achieved by
coupling this term to a chiral multiplet. A similar procedure may be
taken in ${\mathcal N}=2$ supergravity, since there exist ${\mathcal
N}=2$ chiral superfields which must be Lorentz and SU(2) scalars but
can have an arbitrary U(1) weight, allowing for supersymmetric U(1)
breaking couplings.

Such a result should be more difficult to achieve for ${\mathcal N}
\geq 3$, because there are no generic chiral multiplets. But for $3
\leq {\mathcal N} \leq 6$ there are still matter multiplets which
one can couple to the Weyl multiplet. Those couplings could
eventually (but not necessarily) break U(1) $R$-symmetry and lead to
the supersymmetrization of (\ref{r442}).

An even more complicated problem is the ${\mathcal N}=8$
supersymmetrization of (\ref{r442}). The reason is the much more
restrictive character of ${\mathcal N}=8$ supergravity, compared
to lower ${\mathcal N}.$ Besides, its multiplet is unique, which
means there are no extra matter couplings one can take in this
theory. Plus, in this case the $R$-symmetry group is SU(8) and not
U(8): the extra U(1) factor, which in ${\mathcal N}=2$ could be
identified with the remnant ${\mathcal N}=1$ $R$-symmetry and, if
broken, eventually turn the supersymmetrization of (\ref{r442})
possible, does not exist. Apparently there is no way to circunvent
in ${\mathcal N}=8$ the result from \cite{Christensen:1979qj}. In
order to supersymmetrize (\ref{r442}) in this case one should then
explore the different possibilities which were not considered in
\cite{Christensen:1979qj}. Since that article only deals with the
term (\ref{r442}) by itself, one can consider extra couplings to
it and only then try to supersymmetrize. This procedure is very
natural, taking into account the scalar couplings that multiply
(\ref{r442}) in the actions (\ref{2aea4}), (\ref{hea4}).

We now proceed with trying to supersymmetrize (\ref{r442}) but,
first, we review the superspace formulation of ${\mathcal N}\geq
4$ supergravities and also some known higher order superinvariants
in these theories.


\section{Linearized superinvariants in $d=4$ superspace}
\setcounter{equation}{0} \indent

In this section we review the superspace formulation of pure
${\mathcal N}\geq 4$ linearized supergravity theories and some of
the known higher-order superinvariants, including a little
discussion on the symmetries they should preserve. We will only be
working at the linearized level, for simplicity.

One typically decomposes the $\mbox{U}\left({\mathcal N}\right)$
$R-$symmetry into $\mbox{SU}\left({\mathcal N}\right) \otimes
\mbox{U}\left(1\right)$ and considers only $\mbox{SU}\left({\mathcal
N}\right)$ for the superspace geometry. U(1) is still present, but
not in the superspace coordinate indices. The only exception is for
${\mathcal N}=8;$ the more restrictive supersymmetry algebra
requires in this case the $R-$symmetry group to be
$\mbox{SU}\left(8\right),$ and there is no U(1) to begin with. We
always work therefore in this section in conventional extended
superspace with structure group $\mathrm{SL}(2;\mathbb{C}) \otimes
\mbox{SU}\left({\mathcal N}\right).$

\subsection{Linearized ${\cal N} \geq 4, d=4$ supergravity in
superspace} \indent

The field content of ${\mathcal N}\geq 4$ supergravity is
essentially described by a superfield $W^{abcd}$
\cite{Howe:1981gz}, totally antisymmetric in its
$\mbox{SU}\left({\mathcal N}\right)$ indices, its complex
conjugate $\overline{W}_{abcd}$ and their derivatives.

Still at the linearized level, one has the differential relations
\begin{eqnarray}
\nabla_{A a} W^{bcde}&=& -8 \delta_a^{[b} W_A^{cde]}, \nonumber \\
\nabla_{A a} W_B^{bcd}&=& 6 \delta_a^{[b} W_{AB}^{cd]}, \nonumber
\\ \nabla_{A a} W_{BC}^{bc} &=& -4 \delta_a^{[b}
W_{ABC}^{c]}, \nonumber \\ \nabla_{A a} W_{BCD}^b &=& - \delta_a^b
W_{ABCD}, \nonumber \\ \nabla_{A a} W_{BCDE} &=& 0, \label{difw}
\end{eqnarray}
and
\begin{eqnarray}
\nabla_{\dot A}^a W_{BCDE} &=& 2i \nabla_{\underline{B} \dot A}
W_{\underline{C} \underline{D} \underline{E}}^a, \nonumber \\
\nabla_{\dot A}^a W_{BCD}^b &=& i \nabla_{\underline{B} \dot A}
W_{\underline{C} \underline{D}}^{ab}, \nonumber
\\ \nabla_{\dot A}^a W_{BC}^{bc} &=& -i
\nabla_{\underline{B} \dot A} W_{\underline{C}}^{abc}, \nonumber
\\ \nabla_{\dot A}^a W_B^{bcd}&=& i
N^{abcd}_{B \dot A}. \label{difwc}
\end{eqnarray}
This last relation defines the superfield $N^{abcd}_{A \dot A}$
which, therefore, also satisfies
\begin{eqnarray}
N^{abcd}_{A \dot A}&=& \nabla_{A \dot A} W^{abcd}, \\ \nabla_{A a}
N^{bcde}_{B \dot B} &=& -8 \delta_a^{[b} \nabla_{\underline{A} \dot
B} W_{\underline{B}}^{cde]}.
\end{eqnarray}

Here we should notice that these relations are valid for
$N^{abcd}_{A \dot A},$ but not for its complex conjugate
$\overline{N}_{A \dot A abcd}.$ In other words, $\nabla_{A a}
\overline{N}_{B \dot B bcde}$ is another independent relation,
like its hermitian conjugate $\nabla_{\dot A a} N^{bcde}_{B \dot
B},$ as we will see below \cite{Howe:1981gz}.

The spinorial indices in the differential relations (\ref{difw})
are completely symmetrized. Indeed, at the linearized level the
corresponding terms with contracted indices vanish, through the
Bianchi identities
\begin{eqnarray}
\nabla^A_{\, \dot A} W_{ABC}^a &=& 0,
\\ \nabla^A_{\, \dot A} W_{ABCD} &=& 0, \label{dw}
\\ \nabla_{\underline{A}}^{\, \, \dot B}
N^{bcde}_{\underline{B} \dot B} &=&0.
\end{eqnarray}

For ${\cal N} \leq 6,$ $W^{abcd}$ is a complex superfield which
together with $\overline{W}_{abcd}$ describes at $\theta=0$ the $2
\left(
\begin{array}{c} {\cal N} \\4 \end{array} \right)$ real scalars
of the theory. In ${\mathcal N}=8$ supergravity,
the superfield $W^{abcd}$ represents at $\theta=0$ the
$\left(\begin{array}{c} 8
\\4 \end{array} \right)=70$ scalars of the full
nonlinear theory. On-shell, it satisfies the reality condition
\cite{Howe:1981gz}
\begin{equation} W^{abcd} = \frac{1}{4!}\varepsilon^{abcdefgh}
\overline{W}_{efgh}.\label{epsilonw} \end{equation}

Since $N^{abcd}_{A \dot A}= \nabla_{A \dot A} W^{abcd}$, from the
previous relation one also has \emph{on-shell}, in linearized
${\mathcal N}=8$ supergravity,
\begin{equation} N_{A \dot A}^{abcd} = \frac{1}{4!}
\varepsilon^{abcdefgh} \overline{N}_{A \dot A
efgh}.\label{epsilonn}
\end{equation}

Among the derivatives of $W^{abcd}$ there is the superfield
$W_{ABCD}$, which from the differential relations (\ref{difw}) is
related to $W^{abcd}$ at the linearized level by $W_{ABCD} \propto
\nabla_{Aa} \nabla_{Bb} \nabla_{Cc} \nabla_{Dd} W^{abcd} + \ldots$
The Weyl tensor appears as the $\theta=0$ component of $W_{ABCD}$:
\begin{equation}
\left. W_{ABCD} \right| = {\cal W}_{ABCD}. \label{w+}
\end{equation}
Also $\left. W_{BCD}^b \right|$ is the Weyl tensor of the
${\mathcal N}$ gravitinos, $\left. W_{BC}^{bc} \right|$ is the
field strength of $\left(\begin{array}{c} {\mathcal N} \\2
\end{array} \right)$ vector fields and $\left. W_B^{bcd} \right|$
are the $\left(\begin{array}{c} {\mathcal N} \\3 \end{array}
\right)$ Weyl spinors.

In ${\mathcal N}=6, 7$ supergravity there exist extra
$\left(\begin{array}{c} {\mathcal N} \\6 \end{array} \right)$
vector fields, described by $\left. \overline{W}_{BC bcdefg}
\right|.$ In ${\mathcal N}=5, 6, 7$ supergravity there also exist
additional $\left(\begin{array}{c} {\mathcal N} \\5
\end{array} \right)$ Weyl spinors, described by $\left.
\overline{W}_{B bcdef} \right|.$\footnote{In ${\mathcal N}=7$
supergravity there also exists an additional
$\left(\left(\begin{array}{c} {\mathcal N}
\\7
\end{array}\right)=1 \right)$ gravitino. Indeed, the ${\mathcal
N}=7$ and ${\mathcal N}=8$ multiplets are identical.} In
${\mathcal N}=8$ supergravity these superfields do not represent
new physical degrees of freedom, because then we have the
following relations:
\begin{eqnarray}
\overline{W}_{B bcdef} &=& \frac{1}{2} \varepsilon_{bcdefgha}
W_B^{gha}, \nonumber \\ \overline{W}_{BC bcdefg} &=& \frac{1}{6}
\varepsilon_{bcdefgha} W_{BC}^{ha}. \label{epsilonx}
\end{eqnarray}
The differential relations satisfied by these superfields can be
derived, in ${\mathcal N}=8$, from (\ref{epsilonx}) and the
previous relations (\ref{difw}) and (\ref{difwc}). For ${\mathcal
N}\leq 6$ supergravities, which are truncations of ${\mathcal
N}=8,$ these relations are obtained from the ${\mathcal N}=8$
corresponding ones, but considering that (\ref{epsilonw}),
(\ref{epsilonn}) and (\ref{epsilonx}) are not valid anymore (i.e.
by considering $W^{abcd}$ and $\overline{W}_{abcd}$ as independent
superfields). This is the way one can derive the differential
relations which are missing in (\ref{difw}) and (\ref{difwc}),
like $\nabla_{A a} \overline{W}_{bcde}= -\frac{2}{3}
\overline{W}_{A abcde},$ and so on.

Again for $4\leq {\mathcal N}\leq 8$, on-shell (which in
linearized supergravity is equivalent to setting the
$\mbox{SU}\left({\mathcal N}\right)$ curvatures to zero), one has
among others the field equations
\begin{eqnarray}
\nabla^A_{\, \dot A} W_{AB}^{ab} &=& 0,\\ \nabla^{A \dot A}
N^{abcd}_{A \dot A}&=& 0. \label{dn}
\end{eqnarray}
At the component level, at $\theta=0$ (\ref{dn}) represents the
field equation for the scalars in linearized supergravity.
Equations (\ref{epsilonw}), (\ref{epsilonn}) and (\ref{dn}) are
only valid on-shell, and are logically subjected to $\a$
corrections. Plus, most of the equations in this section include
nonlinear terms that we did not include here, but which can be
seen in \cite{Howe:1981gz}.


\subsection{Higher order superinvariants in superspace and their symmetries}
\label{duality} \indent

Next we will be analyzing linearized higher order superinvariants
in superspace.

There are known cases in the recent literature of apparent
linearized ${\mathcal R}^4$ superinvariants in ten-dimensional type
IIB supergravity which did not become true superinvariants
\cite{deHaro:2002vk}. One may therefore wonder if that could not
happen in our case. But in $d=4$ the structure of the transformation
laws and the invariances of the supermultiplets are relatively
easier and better understood than in $d=10$, which guarantees us
that the existence of the full superinvariants from the linearized
ones is not in jeopardy, although they may not fully preserve their
symmetries. We summarize here the explanation which can be found in
\cite{Howe:1980th}.

For ${\mathcal N} \leq 3$, one can get a full nonlinear superspace
invariant from a linearized one simply by inserting a factor of $E,$
the determinant of the supervielbein. This is also true for
${\mathcal N} \geq 4,$ but here some remarks are necessary, as fields
which transform nonlinearly may be present. In these cases, the
classical equations of motion of the theory are invariant under
some global symmetry group $G$. The theory also has a local $H$
invariance, $H$ being the maximal compact subgroup of $G$.
The supergravity multiplet includes a set of
abelian vector fields with a local U(1) invariance. Because of this
invariance, the U(1) potentials corresponding to the vector fields
cannot then transform under $H$ and must be representations of
$G$.

In all these cases in the full nonlinear theory the scalar fields,
represented in superspace by $W^{abcd}$, are elements of the coset
space $G/H$. They do not transform linearly under $G,$ but they
still transform linearly under $H.$ One can use the local $H$
invariance to remove the non-physical degrees of freedom by a
suitable gauge choice. In order for this gauge to be preserved,
nonlinear $G$ transformations must be compensated by a suitable
local $H$ transformation depending on the scalar fields. Because
of this, linearized superinvariants can then indeed be generalized
to the nonlinear case by inserting a factor of $E$, the
determinant of the supervielbein, but they will not have the full
$G$ symmetry of the original equations of motion. If we want the
nonlinear superinvariants to keep this symmetry, we must restrict
ourselves to superfields which also transform linearly, like
those which occur directly in the superspace torsions.

In full nonlinear ${\mathcal N}=8$ supergravity \cite{Brink:1979nt}
$G=\mathrm{E}_{7(7)}$, a real non-compact form of $\mathrm{E}_7$
whose maximal subgroup is $\mbox{SL}(2; \mathbb{R}) \otimes
\mbox{O}(6,6)$ but whose maximal compact subgroup is
$H=\mathrm{SU(8)}$. The 70 scalars are elements of the coset space
$\mathrm{E}_{7(7)}/\mathrm{SU(8)}$. Nonperturbative quantum
corrections break $\mathrm{E}_{7(7)}$ to a discrete subgroup
$\mathrm{E}_7 (\mathbb{Z}),$ which implies breaking the maximal
subgroup $\mbox{SL}(2; \mathbb{R}) \otimes \mbox{O}(6,6)$ to
$\mbox{SL}(2; \mathbb{Z}) \otimes \mbox{O}(6,6; \mathbb{Z}).$
$\mbox{O}(6,6; \mathbb{Z})$ is the $T-$duality group of a
superstring compactified on a six-dimensional torus; $\mbox{SL}(2;
\mathbb{Z})$ extends to the full superstring theory as an
$S-$duality group. In \cite{Hull:1994ys}, evidence is given that
$\mathrm{E}_7 (\mathbb{Z})$ extends to the full superstring theory
as an $U-$duality group. It is this $U-$duality which requires (from
a string theory point of view) that all the 70 scalars of the
${\mathbb T}^6$ compactification of superstring theory are on the
same footing, even if originally, in the $d=10$ theory, the dilaton
is special.

Analogously, for ${\mathcal N}=4$ supergravity coupled to $m$ vector
multiplets, we have $G=\mbox{SL}(2; \mathbb{R}) \otimes
\mbox{O}(6,m)$, $H= \mathrm{U(1)} \otimes \mathrm{O(6)} \otimes
\mathrm{O}(m)$. The conjectured full duality group for the
corresponding toroidally compactified heterotic string, with $m=16$,
is $\mbox{SL}(2; \mathbb{Z}) \otimes \mbox{O}(6,22; \mathbb{Z})$.

The four-dimensional supergravity theories we have been considering
can be seen as low energy effective field theories of toroidal
compactifications of type II or heterotic superstring theories. The
true moduli space of these string theories is the moduli space of
the torus factored out by the discrete T-duality group $\Gamma_T$.
For the case where the left-moving modes of the string are
compactified on a $p$ torus ${\mathbb T}^p$ and the right-moving
modes on a $q$ torus ${\mathbb T}^q$ \cite{Narain:1985jj}, the
moduli space is
$$\left.\frac{\mathrm{SO}(p,q)}{\mathrm{SO}(p) \otimes
\mathrm{SO}(q)}\right/\Gamma_T,$$ with $\Gamma_T= \mathrm{SO}(p,q;
\mathbb{Z})$.

In particular, for type II theories compactified on ${\mathbb T}^6$,
the moduli space is
\begin{equation}
\left.\frac{\mathrm{SO(6,6)}}{\mathrm{SO(6)} \otimes
\mathrm{SO(6)}}\right/\Gamma_T, \label{moduli2}
\end{equation}
with $\Gamma_T= \mathrm{SO(6,6; \mathbb{Z})}$.

For heterotic theories, left-moving modes are compactified on
${\mathbb T}^6$ and right-moving modes on ${\mathbb T}^{22}$,
resulting for the moduli space
$$\left. \frac{\mathrm{SU(1,1)}}{\mathrm{U(1)}} \times \frac{\mathrm{SO(6,22)}}{
\mathrm{SO(6)} \otimes \mathrm{SO(22)}}\right/\Gamma_T,$$ with
$\Gamma_T= \mathrm{SO(6,22; \mathbb{Z})}$. The factor
$\frac{\mathrm{SU(1,1)} }{\mathrm{U(1)}}$ is a separated component
of moduli space spanned by a complex scalar including the dilaton,
which lies in the gravitational multiplet and does not mix with the
other toroidal moduli, lying in the 22 abelian vector multiplets.


\subsection{Some known linearized higher order superinvariants} \indent

In reference \cite{Howe:1981xy}, a general (for all $\mathcal N$)
formalism for constructing four dimensional superinvariants by
integrating over even-dimensional submanifolds of superspace
("superactions") was developed. Using this formalism we will
review some known linearized higher order Riemann superinvariants.
We will mostly be concerned with ${\mathcal N}= 8$
superinvariants, although the results can be easily extended to $4
\leq {\mathcal N} \leq 8$. For a more detailed treatment see
\cite{Howe:1981xy,Howe:2004pn}.

We will start by considering ${\cal W}_+^2 + {\cal W}_-^2,$ the
leading $\a$ correction in the heterotic string effective action.
Its ${\mathcal N}=8$ supersymmetrization at the linearized level is given by
\begin{eqnarray}
&& \a \int \overline{W}_{a_1 a_2 a_3 a_4} W^{a_1 a_2 a_3 a_4}
\mbox{d}^8 \theta  + \mbox{h.c.} \nonumber \\ &\propto& \a
\nabla_{A_1 a_1} \cdots \nabla_{A_4 a_4} \nabla_{b_1}^{A_1} \cdots
\nabla_{b_4}^{A_4} W^{a_1 a_2 a_3 a_4} W^{b_1 b_2 b_3 b_4} +
\mbox{h.c.} \nonumber \\ &=& \a \left(W^{A_1 A_2 A_3 A_4} W_{A_1
A_2 A_3 A_4} + W^{\dot A_1 \dot A_2 \dot A_3 \dot A_4} W_{\dot A_1
\dot A_2 \dot A_3 \dot A_4}\right)\label{w2}.
\end{eqnarray}
Because of the integration measure $\mbox{d}^8 \theta$, (\ref{w2})
is not even an integral over half superspace; yet, this expression
is indeed ${\mathcal N}=8$ supersymmetric (and so are its
${\mathcal N}<8$ truncations). To verify that we recall that at
$\theta=0$ the spinorial superderivatives equal the supersymmetry
transformations: $$\nabla_{A a} \left| = Q_{A a}\right|, \,
\nabla_{\dot A}^a \left|= Q_{\dot A}^a\right|.$$ That $\nabla_{A
a} \nabla^{A_1 a_1} \cdots \nabla^{A_4 a_4} \nabla^{b_1}_{A_1}
\cdots \nabla^{b_4}_{A_4} \overline{W}_{a_1 a_2 a_3 a_4}
\overline{W}_{b_1 b_2 b_3 b_4} =0$ is obvious from the
differential relations (\ref{difw}). From the relations
(\ref{difwc}) one gets after a little algebra
\begin{equation}
\nabla_{\dot B}^b \nabla_{A_1 a_1} \nabla_{A_2 a_2} \nabla_{A_3
a_3} \nabla_{A_4 a_4} W^{a_1 a_2 a_3 a_4} = 2^{10}i \nabla_{A_4
\dot B} \nabla_{A_2 a_2} \nabla_{A_3 a_3} W_{A_1}^{b a_2 a_3}.
\label{qw}
\end{equation}
This way the supersymmetry variation of (\ref{w2}) is proportional to
\begin{eqnarray}
&&\nabla_{\dot B}^b \left[\left(\nabla_{A_1 a_1} \nabla_{A_2 a_2}
\nabla_{A_3 a_3} \nabla_{A_4 a_4} W^{a_1 a_2 a_3 a_4}\right)\left(
\nabla^{A_1}_{b_1} \nabla^{A_2}_{b_2} \nabla^{A_3}_{b_3}
\nabla^{A_4}_{b_4} W^{b_1 b_2 b_3 b_4}\right)\right] \nonumber \\
&=& 2^{11}i \left(\nabla^{A_1}_{b_1} \nabla^{A_2}_{b_2}
\nabla^{A_3}_{b_3} \nabla^{A_4}_{b_4} W^{b_1 b_2 b_3 b_4}\right)
\nabla_{A_4 \dot B} \nabla_{A_2 a_2} \nabla_{A_3 a_3} W_{A_1}^{b
a_2 a_3} \nonumber \\ &=& 4i W^{A_1 A_2 A_3 A_4} \nabla_{A_4 \dot
B} W_{A_1 A_2 A_3}^b =4i \nabla_{A_4 \dot B} W^{A_1 A_2 A_3 A_4}
W_{A_1 A_2 A_3}^b,
\end{eqnarray}
where in the last line we have used (\ref{dw}). This means
(\ref{w2}) is indeed supersymmetric, as it transforms as a
spacetime derivative. We notice that ${\cal W}_+^2 + {\cal W}_-^2$
is, by itself, supersymmetric (the completion is zero). This is no
surprise since, up to non-dynamical Ricci terms, ${\cal W}_+^2 +
{\cal W}_-^2$ is a topological invariant in $d=4$.

The method of \cite{Howe:1981xy} was also used to obtain the
${\mathcal N}=8$ supersymmetrization of ${\cal W}_+^2 {\cal
W}_-^2$ at the linearized level, which is given by
\cite{Kallosh:1980fi}
\begin{eqnarray}
&& \a^3 \int \left(\overline{W}_{a_1 a_2 a_3 a_4} W^{a_1 a_2 a_3
a_4} \right)^2 \mbox{d}^8 \theta \mbox{d}^8 \overline{\theta}
\label{superbr2}
\\ &\propto& \a^3 \int \left[\nabla^{\dot{A}_1 a_1} \cdots
\nabla^{\dot{A}_4 a_4}
\nabla^{b_1}_{\dot{A}_1} \cdots \nabla^{b_4}_{\dot{A}_4}
\overline{W}_{a_1 a_2 a_3 a_4} \overline{W}_{b_1 b_2 b_3 b_4}
\right] \nonumber \\ && \times \left[\nabla_{A_1 c_1} \cdots
\nabla_{A_4 c_4} \nabla^{A_1}_{d_1} \cdots \nabla^{A_4}_{d_4}
W^{c_1 c_2 c_3 c_4} W^{d_1 d_2 d_3 d_4} \right] + \ldots \nonumber
\\ &\propto& \a^3 W^{A_1 A_2 A_3 A_4} W_{A_1 A_2 A_3 A_4} W^{\dot
A_1 \dot A_2 \dot A_3 \dot A_4} W_{\dot A_1 \dot A_2 \dot A_3 \dot
A_4} + \ldots \label{br2}
\end{eqnarray}
The "$\ldots$" represent extra terms at the linearized level
resulting when the dotted and undotted derivatives act together in
the same scalar superfield. Because of all these extra terms the
${\mathcal N}=8$ supersymmetry of (\ref{br2}) is not so obvious,
but it has been shown to be true \cite{Howe:2004pn}.


\section{${\cal W}_+^4 + {\cal W}_-^4$ and extended supersymmetry}
\setcounter{equation}{0} \indent

In this section we turn our attention to the new higher order term
${\cal W}_+^4 + {\cal W}_-^4$ and try to supersymmetrize it at the
linearized level using different methods.

We will only be working at the linearized level, for simplicity.
Therefore we will not be particularly concerned with the string
loop effects considered in the discussion on the string effective
actions , because of their dilaton couplings which are necessarily
highly nonlinear. We will be mainly concerned with the new ${\cal
R}^4$ term in linearized supergravity, not worrying with the
dilatonic factor in front of it to begin with (later this factor
will be considered).


\subsection{Superfield expression of ${\cal
W}_+^4 + {\cal W}_-^4$} \indent

In the same way as ${\cal W}^4=\left({\cal W}_+^2 + {\cal
W}_-^2\right)^2={\cal W}_+^4 + {\cal W}_-^4 +2 {\cal W}_+^2 {\cal
W}_-^2$, the way of writing ${\cal W}^4$ as $\theta=0$ components
of superfields can also be seen - at the linearized level! - as
the "square" of the superfield expression of ${\cal W}^2={\cal
W}_+^2 + {\cal W}_-^2$, given by (\ref{w2}). This way, by "taking
the square" of (\ref{w2}), one obtains (\ref{br2}) and (after
matching the powers of $\a$)
\begin{eqnarray}
&& \a^3 \left[\nabla_{A_1 c_1} \cdots \nabla_{A_4 c_4}
\nabla^{A_1}_{d_1} \cdots \nabla^{A_4}_{d_4} W^{c_1 c_2 c_3 c_4}
W^{d_1 d_2 d_3 d_4} \right]^2 + \mbox{h.c.} \nonumber \\ &\propto&
\a^3 \left(W^{A_1 A_2 A_3 A_4} W_{A_1 A_2 A_3 A_4}\right)^2 + \a^3
\left(W^{\dot A_1 \dot A_2 \dot A_3 \dot A_4} W_{\dot A_1 \dot A_2
\dot A_3 \dot A_4}\right)^2. \label{w4}
\end{eqnarray}
From the differential relations (\ref{difw}), one can see that, at
the linearized level,
\begin{equation}
\nabla_{Aa} \nabla_{Bb} \nabla_{Cc} \nabla_{Dd} \nabla_{Ee}
W^{fghi}=0, \label{d5w}
\end{equation}
from which one sees that, in order for (\ref{w4}) not to vanish,
each $W^{abcd}$ must be acted by four and only four undotted
spinorial derivatives. From (\ref{w4}) one gets a sum of products
of four $W_{ABCD}$ terms. Because of the uniqueness of ${\cal
W}^4$ terms we mentioned, the result must be ${\cal W}_+^4 + {\cal
W}_-^4$.

Therefore, (\ref{w4}) represents the expression of ${\cal W}_+^4 +
{\cal W}_-^4$ in terms of superfields, up to some numerical
factor. The fact that one can write this or any other term as a
superfield component does not necessarily mean that it can be made
supersymmetric; for that one has to show how to get it from a
superspace invariant. In the present case, for (\ref{w4}), the
most obvious candidate for such a superinvariant is
\begin{equation}
\a^3 \int W^{a_1 a_2 a_3 a_4} \overline{W}_{a_1 a_2 a_3 a_4}
W^{b_1 b_2 b_3 b_4} \overline{W}_{b_1 b_2 b_3 b_4} \mbox{d}^{16}
\theta + \mbox{h.c.} \label{w4s}
\end{equation}
By its index structure (it requires sixteen undotted and sixteen
dotted spinorial derivatives), one can see that (\ref{w4s}) is
only valid for ${\mathcal N}=8$ supergravity. But for lower
${\mathcal N}$ an equivalent expression may be written, by
replacing $W^{b_1 b_2 b_3 b_4}$ by some of its spinorial
derivatives, while correspondingly lowering the number of $\theta$
in the measure. For instance, one can obviously write (\ref{w4s})
in an equivalent (at the linearized level) way, which can be more
easily generalized for $4\leq {\mathcal N} \leq 8:$
\begin{equation}
\a^3 \int W^{a_1 a_2 a_3 a_4} \overline{W}_{a_1 a_2 a_3 a_4}
W^{A_1 A_2 A_3 A_4} W_{A_1 A_2 A_3 A_4} \mbox{d}^8 \theta +
\mbox{h.c.} \label{w4s4}
\end{equation}
From (\ref{d5w}), clearly both (\ref{w4s}) and (\ref{w4s4}) are
equivalent to (\ref{w4}) as linearized component expansions (up to
some different numerical factor); now it remains to be seen if
they are indeed supersymmetric. Using (\ref{difw}), (\ref{difwc}),
(\ref{dw}) and (\ref{qw}), one can compute the supersymmetry
variation of the result of the superspace integration, at the
linearized level, which is given by
\begin{eqnarray}
\nabla_{\dot A}^a \left(W^{BCDE} W_{BCDE} \right)^2 &=& -8i
\nabla_{B \dot A} \left(W^{FGHI} W_{FGHI} W^{BCDE} W^a_{CDE}
\right) \nonumber \\ &+& 16i W^{FGHI} \nabla_{B \dot A} \left(
W_{FGHI} W^{BCDE} W^a_{CDE} \right). \label{qw4s4}
\end{eqnarray}
This supersymmetry transformation is not a total derivative and
cannot be transformed into one. Therefore (\ref{w4s}) and its
equivalent (\ref{w4s4}) do not represent a valid superinvariant.
This result is expected: it is just the confirmation of the
prediction from \cite{Christensen:1979qj} in ${\mathcal N}=8$
which, as we said, is not easy to circumvent. The
supersymmetrization of ${\cal W}_+^4 + {\cal W}_-^4,$ if it
exists, must come in a different way.


\subsection{Attempts of supersymmetrization without modification
of the linearized Bianchi identities} \indent

We now try to find out possible ways of supersymmetrizing ${\cal
W}_+^4 + {\cal W}_-^4$ at the linearized level in ${\cal N} \geq
4, d=4$ supergravity in superspace. The known solution to the
superspace Bianchi identities \cite{Howe:1981gz} (equivalent to
the $x$-space supersymmetry transformations) is only valid
on-shell for pure supergravity (without any kind of string
corrections).

In principle, in order to supersymmetrize a higher-order term term
in the lagrangian one needs higher-order corrections to the
superspace Bianchi identities (so one does to the $x$-space
supersymmetry transformation laws), which should be of the same
order in $\a$. In this section we attempt to supersymmetrize
(\ref{r442}) assuming that the solution to the Bianchi identities
for pure supergravity remains valid. This a matter of simplicity:
the complete solution to the Bianchi identities involves, even
without any $\a$ corrections, many nonlinear terms which we
haven't considered \cite{Howe:1981gz}. The $\a$ corrections to the
supersymmetry transformations are necessarily nonlinear and should
affect and generate only nonlinear terms; it does not make sense
to consider them if we are looking only for linearized
superinvariants.

First we check if it is possible to make some change in
(\ref{w4s4}) in order to make it supersymmetric. We notice that
the result in (\ref{qw4s4}) only tells us that (\ref{w4s4}) is not
supersymmetric by itself; it does not mean that it is not part of
some superinvariant. In fact, maybe there exists some counterterm
$\Phi$ which can be added to (\ref{w4s4}) in order to cancel the
supersymmetry variation (\ref{qw4s4}), so that the sum of
(\ref{w4s4}) and $\Phi$ is indeed supersymmetric. In order for
$\Phi$ to exist, it must then satisfy, for some $\Phi_{A \dot A
\dot E }^e,$
\begin{equation}
\nabla_{\dot E}^e \left[\left(W^{ABCD} W_{ABCD}\right)^2 +
\left(W^{\dot A \dot B \dot C \dot D} W_{\dot A \dot B \dot C \dot
D}\right)^2 + \Phi\right]= \nabla^{A \dot A} \Phi_{A \dot A \dot E
}^e.
\end{equation}
Together with (\ref{qw4s4}) this is a very difficult differential
equation, to which we did not find any solution in terms of known
fields, both for $\Phi$ and $\Phi_{A \dot A \dot E}^e.$

The second possibility in order to try to cancel the supersymmetry
variation (\ref{qw4s4}) is to multiply (\ref{w4s4}) by some
factors $\Phi, \overline{\Phi}$, such that the product is
supersymmetric. In this case $\Phi, \overline{\Phi}$ must satisfy,
for some $\Phi_{A \dot A \dot E}^e,$
\begin{equation}
\nabla_{\dot E}^e \left[\overline{\Phi} \left(W^{ABCD}
W_{ABCD}\right)^2 + \Phi \left(W^{\dot A \dot B \dot C \dot D}
W_{\dot A \dot B \dot C \dot D}\right)^2 \right]= \nabla^{A \dot
A} \Phi_{A \dot A \dot E}^e. \label{difw4s4}
\end{equation}
In this case the factors $\Phi, \overline{\Phi}$ must satisfy some
restrictions, both by dimensional analysis (we want an $\a^3$ term)
and by component analysis (we want to supersymmetrize ${\cal W}_+^4
+ {\cal W}_-^4$ in the Einstein frame (\ref{2aea4}) and
(\ref{hea4}), with a factor of $\exp(-4 \phi)$ and at most some
other scalar couplings resulting from the compactification from
$d=10$). Therefore the only acceptable (and actually very natural)
factors $\Phi, \overline{\Phi}$ are simply functions of $W^{abcd},
\overline{W}_{abcd}.$

In any case, again (\ref{difw4s4}) is a very difficult
differential equation, which we tried to solve in terms of each of
the different known fields. We were not able to find any solution,
both for $\Phi, \overline{\Phi}$ and $\Phi_{A \dot A \dot E}^e,$
as one can see by considering (\ref{qw4s4}), which cannot be
canceled simply by taking factors of $W^{abcd},
\overline{W}_{abcd}.$

Therefore one cannot supersymmetrize (\ref{r442}) using only the
linearized (on-shell) solution to the Bianchi identities in pure
supergravity. This result is not so expected and is not a
confirmation of the prediction from \cite{Christensen:1979qj} in
${\mathcal N}=8,$ which applies to (\ref{r442}) by itself and not
when it is multiplied by a scalar factor. In the following
subsection we will use the full nonlinear solution to the Bianchi
identities, but still at $\a=0.$


\subsection{Attempts of supersymmetrization with nonlinear $\a=0$
Bianchi identities} \indent

The generic effective action (\ref{eef}) has a series of terms
which we designated by $I_i({\cal \widetilde{R},M}).$ Some of
these terms can be directly supersymmetrized: they constitute the
"leading terms", each one of them corresponding to an independent
superinvariant. The remaining terms are part of the supersymmetric
completion of the leading ones.

In general it is very hard to determine the number of independent
superinvariants. This problem becomes even more difficult in the
presence of $\a$ correction terms, because one single
superinvariant includes terms at different orders in $\a.$ For the
complete supersymmetrization of a given higher-derivative term of
a certain order in $\a$, typically an infinite series of terms of
arbitrarily high order in $\a$ shows up. This series may be
truncated to the order in $\a$ in which one is working, but when
supersymmetrizing the terms of higher order in $\a$ the
contributions from the lower order terms must be considered. The
reason is, of course, the $\a$ dependence of the supersymmetry
transformations. This has been explicitly shown for (\ref{r441})
and for ${\mathcal N}=1, 2$ in \cite{Moura:2001xx,Moura:2002ip}.
At any given order in $\a,$ therefore, there are new leading terms
(i.e. new superinvariants), and other terms which are part of
superinvariants at the same order and at lower order.

Each time the supersymmetry transformation laws of single fields
include linear terms, it should be possible to determine how to
supersymmetrize an expression written only in terms of these
fields already at the linearized level. A "leading term" of an
independent superinvariant should then be invariant already at the
linearized level. If this linearized supersymmetrization cannot be
found for the term in question, but it still has to be made
supersymmetric, it cannot be a "leading term", and must emerge
only at the nonlinear level, as part of the supersymmetric
completion of some other term. That must be the case of
(\ref{r442}), which we have tried to supersymmetrize directly at
the linearized level and we did not succeed. For the remainder of
this section we will examine that possibility.

Since the $\a$ corrections necessarily introduce nonlinear terms
in the supersymmetry transformations, and since one should not
consider any higher order term before considering all the
corresponding lower order terms, before looking for higher-order
corrections to the supersymmetry transformations one should first
look at their nonlinear $\a=0$ terms. Here we will only be
concerned with the nonlinear terms of the on-shell relations, i.e.
of those relations which will probably acquire $\a$ corrections:
(\ref{epsilonw}), (\ref{epsilonn}) and (\ref{dn}).

The first two linearized equations, (\ref{epsilonw}) and
(\ref{epsilonn}), refer to the 70 scalar fields of ${\mathcal
N}=8$ supergravity. As we mentioned, in the nonlinear theory these
fields are given by the coset space
$\mathrm{E}_{7(7)}/\mathrm{SU(8)}$; they transform nonlinearly
under $\mathrm{E}_{7(7)}$, but they still transform linearly under
SU(8) \cite{Brink:1979nt}. On shell, in superspace, at order
$\a=0$, going from the linearized to the full nonlinear theory
corresponds to replacing the constraint "SU(8) curvature=0" by
"$\mathrm{E}_{7(7)}$ curvature=0". A complete treatment can be
found in \cite{Howe:1981gz}.

The superspace field equation (\ref{dn}) reflects the linearized
field equation of the scalar fields in $4 \leq {\mathcal N} \leq 8$
supergravity, including the dilaton. For the action (\ref{eef}) the
complete dilaton equation is given by
\begin{equation}
\nabla^2 \phi - \frac{1}{2}\ \sum_i \mbox{e}^{\frac{4}{2-d} \left( 1
+ w_i \right) \phi} I_i({\cal \widetilde{R},M}) = 0. \label{bdfe}
\end{equation}
At order $\a=0$, among the terms $I_i({\cal \widetilde{R},M})$
there should be those which contain field strengths corresponding
to each of the vector fields present in the theory. Plus, still at
order $\a=0$ there are couplings of the scalars to fermions, which
we never considered explicitly but must be reflected in their
field equations. In that order in $\a,$ the ${\mathcal N}=8$
nonlinear version of (\ref{dn}), the field equation for the
scalars, is given by \cite{Howe:1981gz}
\begin{eqnarray}
\nabla^{A \dot A} N^{abcd}_{A \dot A} &=& W^{\dot A \dot B}_{ef}
W^{abcdef}_{\dot A \dot B} + 12 W^{AB [ab} W_{AB}^{cd]} \nonumber \\
&+&\frac{i}{12} W^{\dot A}_{efg} W^{A efg} N^{abcd}_{A \dot A}
-\frac{3}{2}i W^{\dot A}_{efg} W^{A e[ab} N^{cd] fg}_{A \dot A}
-\frac{2}{3}i W^{\dot A}_{efg} W^{A [abc} N^{d]efg}_{A \dot A}
\nonumber
\\ &+& 4- \mathrm{fermion \, terms}. \label{dnnl}
\end{eqnarray}
As one can see, this expression does not contain any nonlinear
term which is exclusively dependent on the Weyl tensor. As one can
confirm in \cite{Howe:1981gz}, the same is true for each of the
differential relations considered in (\ref{difw}) and
(\ref{difwc}). Therefore we cannot expect (\ref{r442}) to emerge
from the nonlinear completion of some (necessarily $\a^3$)
linearized superinvariant. One must really understand the
$\a$-corrections to the Bianchi identities. Since these
corrections are necessarily nonlinear, this means one cannot
supersymmetrize (\ref{r442}) at the linearized level at all. Here
one must notice that never happened for the previously known
higher-order terms, which all had its linearized superinvariant.


\subsection{Corrections to the solution of the
linearized Bianchi identities in ${\cal N} \geq 4, d=4$
superspace: some considerations} \indent

In each of the three effective actions (\ref{2bea4}),
(\ref{2aea4}), (\ref{hea4}), only the ${\cal W}_+^2 {\cal W}_-^2$
term contains the transcendental coefficient $\zeta(3)$. This term
must then have its own superinvariant, as no other term has such a
coefficient. Therefore the changes in the supersymmetry
transformation laws the other terms generate do not have such a
coefficient and could not, by themselves, cancel the supersymmetry
variation of  (\ref{r441}).

Since the numerical coefficient in front of (\ref{r442}) in the
$d=4$ effective actions (\ref{2aea4}) and (\ref{hea4}) is not
transcendental, this term may eventually not need its own
superinvariant and be part of some other superinvariant, with a
different leading bosonic term, maybe even of a lower order in
$\a$, being related to (\ref{r442}) by an $\a$-dependent
supersymmetry transformation. But even if such relation is valid
in $d=4$, that does not mean at all it should keep being valid in
$d=10$.

One can try to generate a higher-order (in $\a$) term from a
lower-order higher derivative superinvariant; maybe the
higher-order term would lie on the orbit of its supersymmetry
transformations. But in order to generate the higher-order term
this way, one obviously needs to know the $\a$-corrected
supersymmetry transformation laws.

One possibility would be to see if (\ref{r442}) could be obtained
from the supersymmetrization of the ${\cal W}^2$ term in
(\ref{w2}), of order $\a$. But this term does not come from type
II theories, which only admit $\a^3$ corrections and higher; it
only comes from the heterotic theories. Therefore a ${\cal W}^2$
term must only be present as a correction to ${\mathcal N}=4$
supergravity: it can also be written as an ${\mathcal N}=8$
invariant, given by (\ref{w2}), but in this case its stringy
origin is not so obvious. Indeed, ${\cal R}^2$ terms show up from
the ${\cal R}^4$ terms we are considering when we compactify
string theory on a Calabi-Yau manifold \cite{Antoniadis:1997eg},
but for the moment we are only considering toroidal
compactifications with maximal $d=4$ supersymmetry.

There are other different terms one can consider. For instance,
when going from the string frame (\ref{esf}) to the Einstein frame
(\ref{eef}) with the transformation (\ref{rsre}), one gets from a
polynomial of the Riemann tensor a dilaton coupling and powers of
derivatives of $\phi$. The $\a^3$ effective action should contain,
besides (\ref{r441}) and (\ref{r442}), the terms
$\left(\left(\nabla^{\mu} \nabla^{\nu} \phi \right)
\left(\nabla_{\mu} \nabla_{\nu} \phi \right) \right)^2,$
$\left(\nabla^{\mu} \nabla^{\nu} \phi \right) \left(\nabla_{\mu}
\nabla_{\nu} \phi \right) \left(\nabla^2 \phi \right)^2$ and
$\left(\nabla^2 \phi \right)^4.$

Taking as an example the $\a^3$ term $\left(\nabla^2 \phi
\right)^4$, it can be represented in superspace as part of $\left.
\left[\left(\nabla^{A \dot A} N^{abcd}_{A \dot A}\right)
\left(\nabla_{B \dot B} \overline{N}_{abcd}^{B \dot
B}\right)\right]^2 \right|,$ which can indeed be supersymmetrized:
from (\ref{difw}) and (\ref{difwc}), this term should come from
(\ref{superbr2}) by acting in each $W^{abcd}$ with two undotted
and two dotted spinorial derivatives (the same for
$\overline{W}_{abcd}$). This should then be one of the terms
represented by the dots in (\ref{br2}).

One therefore may expect the supersymmetrization of the higher
derivative term $I({\cal R})$ (which in the case we are interested
includes ${\cal W}_+^4 + {\cal W}_-^4$) to lie in the orbit of
some power of $\nabla^2 \phi$ or some other superinvariant of
lower order in $\a$, so that one term may result from the other
via an $\a$ dependent supersymmetry transformation. If that is the
case, one needs to find the $\a$ corrections to the (on-shell)
solution of the superspace Bianchi identities, namely to the
nonlinear versions of (\ref{epsilonw}), (\ref{epsilonn}) and
especially (\ref{dn}).

Let's take for example the nonlinear dilaton field equation.
Considering the pure gravitational $\a$ corrections expressed in
the effective actions (\ref{2bea4}), (\ref{2aea4}), (\ref{hea4}),
we are able to "guess" the expected corrections to (\ref{dnnl}),
knowing the field content of $W^{abcd}$ and its derivatives.
Neglecting for the moment the numerical coefficients, one can see
that some of the expected corrections to (\ref{dnnl}) (only the
purely gravitational ones, i.e. those depending only on the Weyl
tensor) are of the form
\begin{eqnarray}
\left. \nabla^{A \dot A} N^{abcd}_{A \dot A} \right|_{\a + \a^3}
&\propto& \a W^{abcd} \left[W^{ABCD} W_{ABCD} + W^{\dot A \dot B
\dot C \dot D} W_{\dot A \dot B \dot C \dot D} \right]\nonumber
\\ &+&\a^3 W^{abcd} \left[\left( W^{ABCD} W_{ABCD}\right)^2 +
\left( W^{\dot A \dot B \dot C \dot D} W_{\dot A \dot B \dot C
\dot D} \right)^2 \right. \nonumber \\ &+& \left. \left( W^{ABCD}
W_{ABCD}\right)\left( W^{\dot A \dot B \dot C \dot D} W_{\dot A
\dot B \dot C \dot D} \right) \right] + \ldots \label{dna}
\end{eqnarray}
Of course this equation must be completed with other
contributions, which may be derived, including the numerical
coefficients, (\ref{2aea4}) and (\ref{hea4}), once they are
completed with the other leading $\a$ corrections which do not
depend only on the Riemann tensor.

It remains to be seen how are these corrections compatible with
the superspace Bianchi identities. This would allow us to
determine the $\a$ corrections one needs to introduce in the other
superspace field equations in order to the superspace Bianchi
identities remain valid to this order in $\a$. This is a
technically very complicated problem which we are not addressing
in the present work.

\subsection{${\cal W}_+^4 + {\cal W}_-^4$, $U-$duality and ${\mathcal N}=8$
supergravity} \label{uduality}\indent

As we mentioned before, the "no-go theorem" for the
supersymmetrization of (\ref{r442}) given in
\cite{Christensen:1979qj} is based on ${\mathcal N}=1$ chirality
arguments. In order to circumvent these arguments, a reasonable
possibility is to try to construct a superinvariant which violates
the U(1) symmetry or (for ${\mathcal N}>1$) some of the
$R$-symmetry. But the superfield expression corresponding to
(\ref{r442}) given by (\ref{w4}) is even U(1)-symmetric, as
$W_{ABCD}$ is U(1)-invariant. (This is more clearly derived in
${\mathcal N}=1$ superspace \cite{Moura:2007ks}, but it is easily
understood if one thinks that from (\ref{w+}) $\left. W_{ABCD}
\right|$ is a component of the Riemann tensor.) The best one can
aim at is to break U(1) or part of the $\mbox{SU}\left({\mathcal
N}\right)$ by taking a different integration measure, as suggested
in \cite{Howe:1981xy} and as we tried with (\ref{w4s4}). In
${\mathcal N}=8$ superspace one can keep trying extra couplings of
the scalar superfields $W^{abcd}$ combined with different
nonstandard integration measures. But it is easier if we are
allowed to consider other multiplets than the gravitational, whose
couplings automatically violate U(1). That is not possible in
${\mathcal N}=8$ supergravity, both because there are no other
multiplets than the gravitational to consider, and because the
extra U(1) symmetry does not exist. We recall that ${\mathcal N}
\leq 6$ theories have a $\mbox{U}\left({\mathcal N}\right)$
symmetry, which is split into $\mbox{SU}\left({\mathcal N}\right)
\otimes \mbox{U}\left(1\right)$, but the more restrictive
${\mathcal N}=8$ theory has originally only an SU(8) symmetry.
This may be part of the origin of all the difficulties we faced
when trying to supersymmetrize (\ref{r442}) in ${\mathcal N}=8$.

But the main obstruction to this supersymmetrization is that,
opposite to ${\cal W}_+^2 {\cal W}_-^2,$ the term ${\cal W}_+^4 +
{\cal W}_-^4$ does not seem to be compatible with the full
$R-$symmetry group SU(8). In ref. \cite{Drummond:2003ex}, a complete
study has been made of all possible higher-order terms in
${\mathcal N}=8$ supergravity, necessarily compatible with SU(8),
and (\ref{r442}) does not appear in the list of possible terms.

Indeed, as we saw in the discussion of section \ref{duality}, only
the local symmetry group of the moduli space of compactified
string theories should be preserved by the four dimensional
perturbative string corrections. As we saw in (\ref{moduli2}), for
${\mathbb T}^6$ compactifications of type II superstrings this
group is given by $\mbox{SO}(6) \otimes \mbox{SO}(6) \sim
\mbox{SU}(4) \otimes \mbox{SU}(4)$, which is a subgroup of SU(8).
Most probably the perturbative string correction term ${\cal
W}_+^4 + {\cal W}_-^4$ only has this $\mbox{SU}(4) \otimes
\mbox{SU}(4)$ symmetry. If that is the case, in order to
supersymmetrize this term besides the supergravity multiplet one
must also consider $U-$duality multiplets \cite{Bars:1995bv}, with
massive string states and nonperturbative states. These would be
the contributions we were missing.

But in conventional extended superspace one cannot simply write
down a superinvariant that does not preserve the
$\mbox{SU}\left({\mathcal N}\right)$ $R-$symmetry, which is part
of the structure group. One can only consider higher order
corrections to the Bianchi identities which preserve
$\mbox{SU}\left({\mathcal N}\right)$, like the ones from
(\ref{dna}), but these corrections would not be able to
supersymmetrize (\ref{r442}). ${\mathcal N}=8$ supersymmetrization
of this term would then be impossible; the only possible
supersymmetrizations would be at lower $\mathcal N$, eventually
consider $U-$duality multiplets.

The fact that one cannot supersymmetrize in ${\mathcal N}=8$ a
term which string theory requires to be supersymmetric, together
with the fact that one needs to consider nonperturbative states
(from $U-$duality multiplets) in order to understand a
perturbative contribution may be seen as indirect evidence that
${\mathcal N}=8$ supergravity is indeed in the swampland, as
proposed in \cite{Green:2007wx}. We believe that topic deserves
further study.

\section{Conclusions}\setcounter{equation}{0}
\indent

We had shown in \cite{Moura:2007ks} that type IIA and heterotic
string theories predict the term ${\cal W}_+^4 + {\cal W}_-^4$ to
show up at one loop when compactified to four dimensions.
Nonetheless, an older article \cite{Christensen:1979qj} stated
that this term, by itself, simply could not be made supersymmetric
in $d=4$. In \cite{Moura:2007ks} we worked out its ${\mathcal
N}=1$ supersymmetrization, by coupling it to a chiral multiplet.
In this article we considered the more complicated problem of its
${\mathcal N}=8$ supersymmetrization. We obtained the superfield
expression of that term, given by (\ref{w4}), and we have shown
that expression indeed was not part of a superinvariant.

Since that term in $d=10$ should come coupled to a dilaton, and it
may acquire other scalar couplings after compactification to
$d=4$, in order to try to circumvent the argument of
\cite{Christensen:1979qj} we tried to construct a superinvariant
which included this term, together with a proper scalar coupling,
in general $4 \leq {\mathcal N} \leq 8$ superspace. We concluded
that the supersymmetrization of this term at the linearized level,
by itself, cannot be achieved, something which was always possible
for the previously known higher-derivative string corrections.

We proposed some changes to the on-shell solution to the
superspace Bianchi identities in order to include the lowest order
$\a$-corrections. We did not present the whole set of possible
$\a$-corrections to the Bianchi identities nor we tried to solve
them in order to check the consistency of these corrections and to
determine their coefficients. In ${\mathcal N}=8$ superspace one
can only have SU(8) invariant terms, and we argued ${\cal W}_+^4 +
{\cal W}_-^4$ should be only $\mbox{SU}(4) \otimes \mbox{SU}(4)$
invariant. If that is the case, in order to supersymmetrize this
term besides the supergravity multiplet one must introduce
$U-$duality multiplets, with massive string states and
nonperturbative states. ${\mathcal N}=8$ supersymmetrization of
(\ref{r442}) may not be possible at all, which may be another
argument favoring the hypothesis that ${\mathcal N}=8$
supergravity is in the swampland \cite{Green:2007wx}. This is a
very fundamental topic of study, together with the recent claims
of possible finiteness of ${\mathcal N}=8$ supergravity. Plus, as
we concluded from our analysis of the dimensional reduction of
order $\a^3$ gravitational effective actions, the new ${\cal R}^4$
term (\ref{r442}) has its origin in the dimensional reduction of
the corresponding term in M-theory, a theory of which there is
still a lot to be understood. We believe therefore that the
complete study of this term and its supersymmetrization deserves
further attention in the future.


\section*{Acknowledgments} \indent

I wish to thank Pierre Vanhove for very important discussions and
suggestions which made it possible to arrive at the results of
section \ref{uduality}, and for useful comments on the manuscript.
I also wish to thank Paul Howe for discussions and Radu Roiban for
correspondence. It is a pleasure to acknowledge the excellent
hospitality of the Service de Physique Th\'eorique of CEA/Saclay
in Orme des Merisiers, France, where some parts of this work were
completed.

This work has been supported by Funda\c c\~ao para a Ci\^encia e a
Tecnologia through fellowship BPD/14064/2003 and Centro de L\'ogica
e Computa\c c\~ao (CLC).


\appendix


\section{Superspace conventions}
\label{appendix1} \setcounter{equation}{0} \indent

The superspace conventions for index manipulations and complex
conjugations are essentially the same as in \cite{Moura:2002ip}.
Underlined (resp. in square brackets) indices are symmetrized
(resp. antisymmetrized) with weight one, i.e.
$$X_{\underline{AB}}= \frac{X_{AB} + X_{BA}}{2}, \,\, X_{[ab]}=
\frac{X_{ab}- X_{ba}}{2}.$$

At the linearized level, when interchanging superspace covariant
derivatives, we take all the supertorsions/curvatures to zero with
the exception of \begin{equation} T_{A a \dot B}^{\, \, \, \, \,
\, \, \, b m}=-2i \delta^b_a \sigma_{A \dot B}^m.
\label{tm}\end{equation} For a complete treatment of superspace
supergravity at the nonlinear level, including the solution to the
superspace Bianchi identities, we refer the reader to
\cite{Howe:1981gz}. In the paper we just summarize the results we
need.


\end{document}